# Hybrid integrated narrow linewidth semiconductor laser based on the distributed feedback from an external deformed microcavity

Da Wei, Leilei Shi*, Yujia Li, Minzhi Xu, Chaoze Zhang, Xianming Huang, Jianxian Yu, Lei Zhai, Wenxuan Huang, Huan Tian, And Tao Zhu**

*Key Laboratory of Optoelectronic Technology & Systems (Education Ministry of China), Chongqing University, Chongqing 400044, China*

*shileilei@cqu.edu.cn; zhutao@cqu.edu.cn*

**Abstract**

Optical microcavities with rotational symmetry have been widely used for narrowing linewidth and reducing frequency noise, however, the narrow but wavelength dependent optical feedback restricts the narrow linewidth laser works only at some discrete wavelength matching the resonance of the microcavity. Here, we demonstrate a narrow linewidth semiconductor laser with continuous wavelength tunability by hybrid integrating a DFB laser chip with a deformed microcavity fabricated on a 220 nm SOI wafer. The deformed microcavity with vortex radius demonstrates the unique characteristics of unidirectional energy storage, wavelength self-adaptivity, and self-focusing of the Rayleigh scattering based distributed feedback. In addition, the strength of Rayleigh scattering is also significantly enhanced by the high numerical aperture silicon waveguide. The optical feedback signal measured by the optical frequency domain reflectometry (OFDR) shows that the deformed microcavity can effectively lengthen the equivalent propagation distance without wavelength dependence. With the wavelength self-adaptive optical feedback from the deformed microcavity, the intrinsic linewidth of a DFB laser diode is narrowed to 525 Hz and the side mode suppression ratio (SMSR) is improved to 76 dB in a maximum allowable continuous wavelength tuning range of 2.25 nm. The frequency noise and relative intensity noise (RIN) are reduced to 2.98 $Hz^2/Hz$ and -148.74 dB/Hz at the offset frequency of 1 MHz, respectively. The work demonstrated here paves a new way for integrated tunable narrow linewidth lasers, which are of crucial importance in high-speed communication and high-precision spectroscopy.

**Introduction**

Lasers with narrow linewidth and low noise are in high demand in both fundamental researches and engineering applications. For instance, atomic clocks require lasers with hertz-level linewidth to probe and manipulate atomic transitions[1-3]. The coherent communication over hundreds of kilometers[4,5] and LIOG gravitational wave detection[6,7] necessitates lasers with a linewidth of kilohertz or less. In addition, a narrow linewidth laser with low noise also plays an crucial role in gyroscope[8,9], LiDAR[10,11], spectroscopy[12], optical frequency synthesis[13,14], microwave photonics[15-17] and other promising applications. Compared to the commonly used solid-state lasers and fiber lasers, semiconductor lasers demonstrate unparalleled advantages in size, weight, power consumption, and cost. Nevertheless, the linewidth of commercially available semiconductor lasers is typically on the order of hundreds of kilohertz or even a few megahertz, which is several orders of magnitude higher than the bulk fiber lasers.

External cavity with high quality-factor ($Q$-factor) has been proven a powerful tool for linewidth narrowing which provides optical feedback by Rayleigh backscattering from an all-pass microresonator[18-21] or the resonant feedback from an add-drop microresonator[22-24]. The linewidth narrowing ratio scales quadratically with the $Q$-factor but requires that the lasing wavelength matches the resonance of the external cavity[18,25,26]. Thermal tuning or current regulation of the III-V chip and localized temperature control, electro-optic effect or free-carrier dispersion effect of the external cavity have been utilized to achieve the match between the lasing chip and the external cavity[27-30]. It is foreseeable that when lasers operate in multi-wavelength arrayed configurations, the tuning complexity and the system redundancy will explode exponentially with the increase in the number of lasing wavelengths, which is detrimental to the integration of the system.

The ubiquitous Rayleigh scattering in a wavelength-independent waveguide provides distributed weak optical feedback with the characteristics of continuous phase and wavelength self-adaptivity[31,32]. Such distributed weak feedback narrows the linewidth by introducing spatiotemporal perturbations to the standing wave within the main laser cavity, exciting the residual inverted population in the upper energy level, and finally reducing the spontaneous emission noise. Benefitting from the continuous phase distribution, the feedback signal can automatically match the phase of the lasing field, significantly simplifying the system by removing the tuning process and controlling modules[33-35]. Currently，Rayleigh scattering in a high-Q microresonator can only be accumulated at the specific resonant wavelength[18-21]. While the accumulation of Rayleigh scattering in a waveguide without wavelength selectivity is achieved by lengthening it to tens or even hundreds of meters[36-38].

In this context, we introduce an on-chip deformed microcavity with vortex radius as the external cavity to provide distributed weak feedback for hybrid integrated narrow linewidth semiconductor laser. By leveraging the deformed microcavity with characteristics of unidirectional energy storage, high backscattering excitation efficiency, wavelength self-adaptivity, and self-focusing of the backscattered light, we achieved an intrinsic linewidth of 525 Hz in a maximum allowable wavelength tuning range of 2.25 nm. The white frequency noise is reduced to the level of 2.98 $Hz^2/Hz$ beyond the offset frequency of 1 MHz. The RIN of the hybrid integrated laser is -148.74 dB/Hz@1 MHz, which is 2 orders of magnitude lower than the free-running laser. The mechanism for efficiently accumulating Rayleigh backscattering pave a new way not only for integrated tunable narrow linewidth lasers but also for the integration of random lasers and chaotic lasers.

## Deformed microcavity concept, theory, simulation, fabrication, and characterization

The schematic diagram of the hybrid integrated narrow linewidth laser with deformed microcavity is shown in Fig.1. Half of the light from the original DFB lasing cavity injects into the deformed microcavity after passing through a multi-mode interferometer (MMI), exciting Rayleigh scattering signal as it travels along the deformed microcavity. The backward Rayleigh scattering signal is fed back into the lasing cavity to perturb the phase of the longitudinal mode, by which more inversion population can be used for stimulated emission, thus significantly reducing the noise resulted from the spontaneous emission and narrowing the linewidth. In addition to integration, the deformed microcavity also demonstrate advantages of thermal and acoustical stabilities over the commonly used optical fiber providing Rayleigh scattering based distributed feedback.

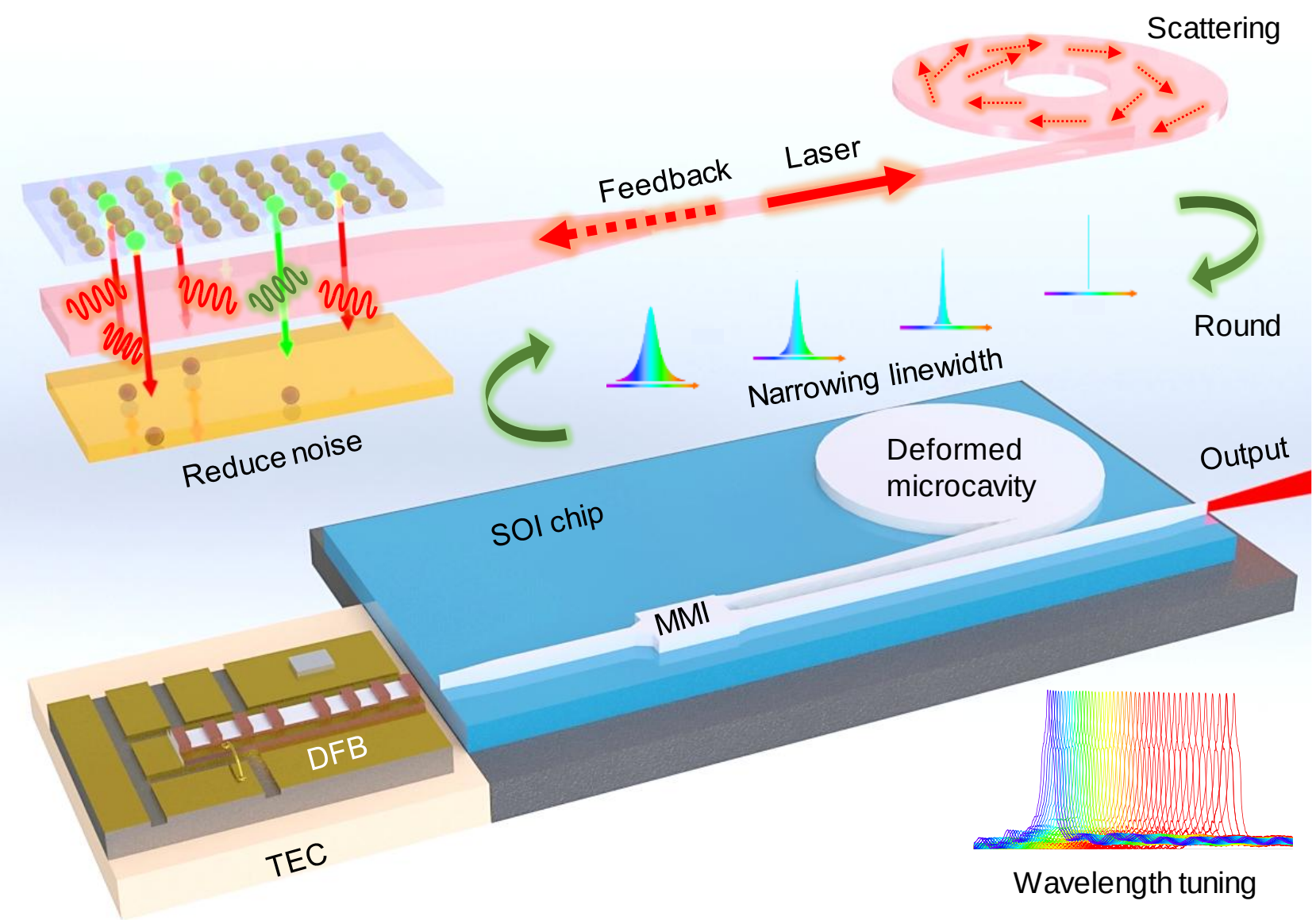

Fig. 1. Deformed microcavity and hybrid integrated semiconductor laser concept

Fig.1. Schematic diagram of hybrid integrated narrow linewidth laser with an external deformed microcavity

The deformed microcavity with vortex radius is schematically shown in Fig.2a. The radius change follows a vortex function: $R=R_{max}+c\varphi$, where $R$ is the cavity radius, $R_{max}$ is the initial or maximum radius, $c=\Delta R/2\pi$ is the radius variation rate, $\Delta R=R_{max}-R_{min}$ is the radius difference, $R_{min}$ is the minimum radius, and $\varphi$ is the circumferential angle. A mode-field converter is used to lower the mode field mismatch induced loss between the lasing cavity, resulting in a notch to the deformed microcavity. The widths of the 100 μm-long linearly tapered mode-field converter are 0.45 μm and $\Delta R$, respectively. As the light enters into the deformed microcavity, backward light occurs accompanying the forward light due to Rayleigh scattering. For the forward light, the incident angle at the microcavity-cladding interface progressively decreases until it falls below the critical angle for total internal reflection since the curvature radius of the cavity gradually decreases. For the backward light, i.e., Rayleigh backscattered signal, the incident angle progressively increases with the gradually increased cavity curvature radius, protecting it from leaking out the deformed microcavity and guiding it focus towards the mode converter. Therefore, as long as the initial reflection angle exceeds the critical angle for total internal reflection, scattered rays propagating in different directions will progressively converge toward the converter after undergoing sufficient rounds of total internal reflection within the cavity. The distinct behaviors of forward and backward light in the deformed microcavity result in intriguing characteristics of forward light storage and backward light self-collection, which facilitate efficient excitation and collection of Rayleigh scattering signal. Furthermore, light circulates along different paths in the deformed microcavity due to its asymmetry. Consequently, the deformed microcavity demonstrates no wavelength dependence, i.e., wavelength self-adaptivity.

By setting $\alpha=\Delta R/R_{max}$ and varying it from 0.0018 to 0.036, we calculated the Poincaré surface of section (PSOS) of the deformed microcavity, as shown in Fig. 2b (more detailed results in Fig. S2). The angles $\varphi$ and $\theta$ represent the azimuthal angle where total internal reflection occurs and the angle between the direction of light movement and the normal of the boundary[39,40]. Light leakage occurs when the reflection angle falls below the critical angle, locating in the run-away area. The size of the run-away area is determined by the critical angle, which itself depends on the refractive index difference between the deformed microcavity and its surrounding medium. The energy of the light located in the run-away area decays rapidly. Therefore, we only statistically analyze the total internal reflection events hereafter. In Fig. 2b, we connected the scatter points in the PSOS for $\alpha$=0.036 according to the propagation path (i.e., the chronological sequence of total internal reflection events). It can be clearly observed that as light propagates within the cavity, the phase $\varphi$ changes by $2\pi$ (from $-\pi$ to $\pi$) per round trip, while the angle $\theta$ decreases linearly with the increase of $\varphi$. After completing one roundtrip, the scatter points in the PSOS jump to the next row (note that such row-skipping does not occur in resonant cavities, whereas in chaotic cavities, occurs randomly[39-44]), corresponding to the subsequent roundtrip. Based on statistical analysis, the deformed cavity with $\alpha$=0.036 exhibits a total of 16 rows in its PSOS, indicating that light undergoes 16 roundtrips via total internal reflection within the cavity. The number of circulation cycles increases significantly as $\alpha$ decreases. Figure 1e shows the statistic of the angle $\theta$ according to the sequence of total internal reflection events (i.e., the number of total reflections), indicating that the angle $\theta$ decreases monotonically with the number of reflections. For each round trip, corresponding to a $2\pi$ change in $\theta$, a step-like jump appears due to the notch (inset in Fig. 2e).

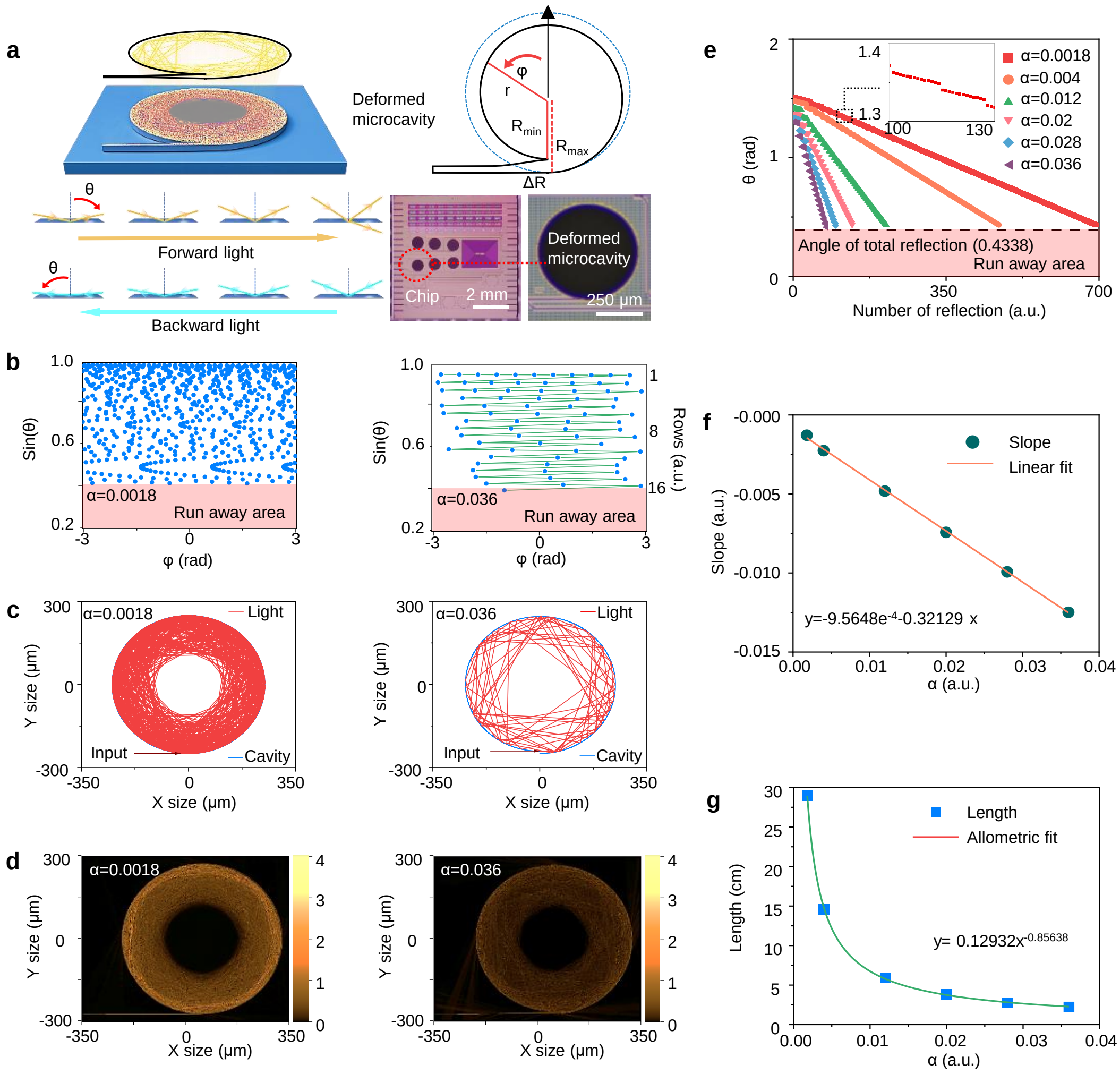

**Fig. 2. Deformed microcavity structure and simulation analysis**. **a.** Schematic diagram and the micrograph of the deformed microcavity. **b.** PSOS of the deformed microcavity. **c.** Numerically calculated propagation path inside the deformed microcavity. **d.** Simulated electric field distribution inside the deformed microcavity. **e.** Statistic of the total internal reflection angles of the deformed microcavity with different α. **f.** Dependence of the variation of the reflection angle on the parameter α. **g.** Equivalent propagation length of the total internal reflected light within the deformed microcavity. The complete simulation data in (b), (c), and (d) are shown in the supplementary material, and the data shown here are only for the cases where α are 0.0018 and 0.036.

The monotonic linear change in the $\theta$ indicates that the light propagates non-resonantly within the cavity. However, the propagation distance and the total internal reflection path in the proposed deformed microcavity can be precisely controlled by $\alpha$, which is completely different from the chaotic microcavity with random $\theta$. The propagation path in the deformed microcavity with different $\alpha$ is shown in Fig. 2c, indicating that the number of total internal reflection decreases with $\alpha$. The propagation distance inversely decreases with $\alpha$, as shown in Fig.2g. It is worth noting that the input light at different wavelength demonstrate the same propagation behavior due to the non-resonance of the deformed microcavity. In addition to the numerical calculation, we also conducted simulation about the field distribution of the deformed microcavity using the finite-difference time-domain (FDTD) method, as shown in Fig.2d. The initial radius of the deformed microcavity is 250 μm in the simulation, which is the same as that used for numerical calculation. The incensement of $\alpha$ decreases the intracavity energy density by lowering the total energy and enlarging the mode field volume, which agrees well with the numerical simulation.

## Experimental results and discussion

According to the theoretical analysis, the critical angle also plays a crucial role in the energy storage and Rayleigh backscattering collection in addition to the structural parameter $\alpha$. Silicon demonstrates the highest refractive index around 1550 nm among the currently available materials used for photonic integrated circuit, by which the smallest run-away area in the PSOS can be achieved. Moreover, the high numerical aperture of the SOI waveguide enables the high-efficiency collection of Rayleigh scattering[45,46], which results from the large refractive index difference between the silicon core and silica cladding. Based on the benefits for light storage and Rayleigh scattering collection, we chose SOI as the platform to fabricate the deformed microcavity.

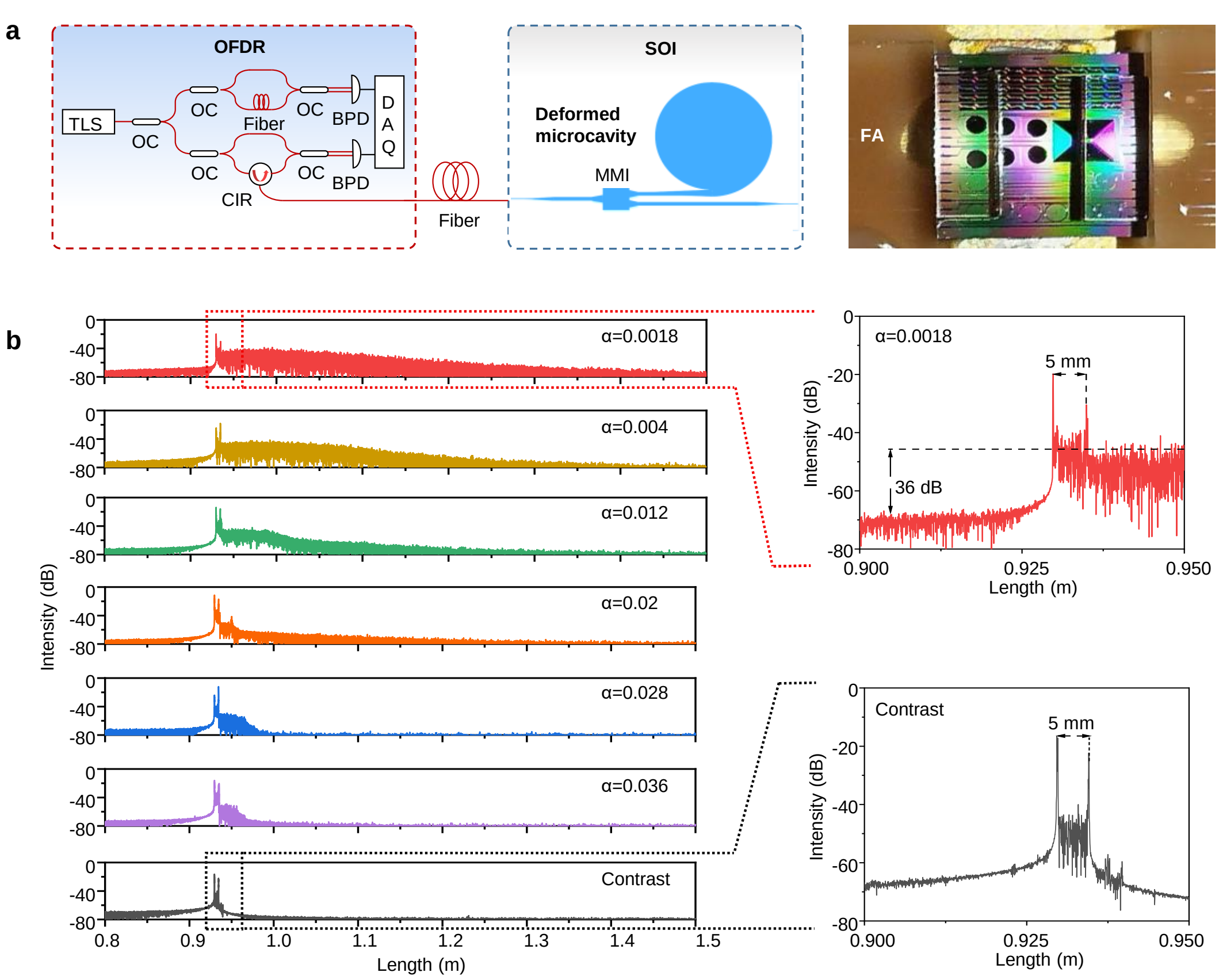

**Fig. 3. Performance of the deformed microcavity. a.** Experimental setup used for measuring the backscattering signal and a micrograph of the FA-coupled chip. **b.** Backscattering signal of the deformed microcavities with different α, where the lowest panel shows the case without deformed microcavity for comparison. A zoom-in view is provided for α = 0.0018 and the "Contrast" backscattering intensity . The backscattering peaks at both ends correspond to reflections from the chip facets, with a peak spacing of ~5 mm matching the chip length.

To characterize Rayleigh backscattering of the deformed microcavity with an optical frequency domain reflectometry (OFDR)[47-50], we couple the light into the SOI chip using a lensed fiber array, as shown in Fig.3a. The light is equally divided into two parts by a MMI after passing through a tapered edge coupler (EC) with a mode field diameter of 3.5 μm. One enters into the deformed microcavity via a mode converter, which excite backward light due to Rayleigh scattering. The other passing through another tapered EC is used for laser characterization. Fig. 3b shows the measured backscattering signals from the deformed microcavities with $\alpha$ varying from 0.0018 to 0.036. For comparison, we chose a 5-mm-long straight single-mode waveguide as the reference, the length of which equals to the sum of ECs, MMI and the deformed microcavity. It can be found from the experimental results that the equivalent propagation length generating Rayleigh scattering signal decreases gradually with $\alpha$, which agrees well with the theoretical analysis. The backscattering intensity of the deformed microcavity with a $\alpha$ of 0.0018 is approximately ~36 dB higher than that of a standard single-mode fiber, as shown in Fig. 3b.

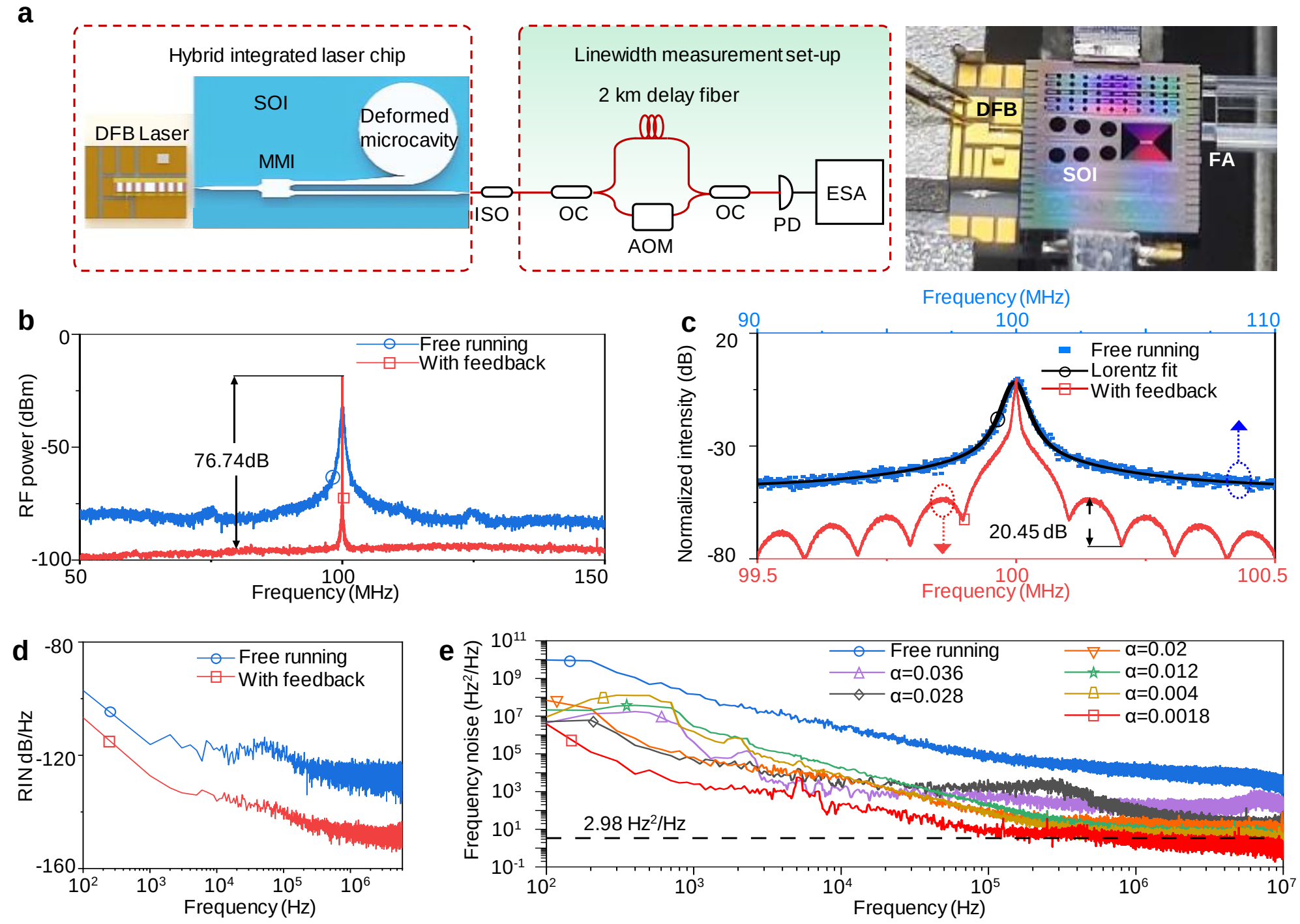

**Fig. 4. Performance of the hybrid integrated narrow linewidth laser . a** Schematic, micrograph of the hybrid integrated laser. **b** SMSRs of the laser with and without distributed feedback. **c** Normalized coherent spectra of laser with and without distributed feedback measured via short-delay DSHI. The linewidth can be derived from the second peak-to-trough contrast of the coherence envelope. **d** RIN of the laser with and without distributed feedback. **e** Laser frequency noise with distributed feedback from deformed microcavities with different α.

The schematic diagram and micrograph of the hybrid integrated semiconductor laser are shown in Fig.4a. The output from a III-V DFB laser is edge coupled to the SOI chip and characterized by a short-delayed self-heterodyne interferometry (DSHI). The short delayed DSHI consists of an unbalanced Mach-Zehnder interferometer (MZI), a photodetector (PD, THORLABS PDB450C), and a spectrum analyzer (ESA, ROHDE & SCHWARZ FSV Signal Analyzer, 10 Hz~30 GHz). The two arms of the MZI are equipped with an acousto-optic modulator (AOM) with a frequency shift of 100 MHz and a 2 km-delay fiber, respectively. The electrical side mode suppression ratio (SMSR) of the DFB laser diode with external distributed feedback provided by the deformed microcavity is 76.74 dB, as shown in Fig. 4b, which is 38.10 dB higher than the laser diode under the free-running state. Clear coherent envelope can be observed by zooming-in the beat spectrum in a frequency range of 1 MHz, as shown in Fig. 4d. According to the method in [51,52], the 20.45 dB contrast between the second peak and trough corresponds to a Lorentzian linewidth of 525 Hz. While the free-running DFB laser diode demonstrate a Lorentzian linewidth of 121.23 kHz without interference envelope. The RIN of the laser diode with external distributed feedback is reduced to -148.74 dB/Hz beyond the offset frequency of 1 MHz, which is 29.12 dB lower than the laser diode under the free-running state. Fig. 4g shows the frequency noise spectra of the laser diode with different external distributed feedback, indicating that smaller deformation results in more frequency noise reduction. Specifically, as the deformation parameter of the microcavity $\alpha$ =0.0018, the frequency noise of the DFB laser diode is reduced to the level of 2.98 $Hz^2/Hz$ at the offset frequency of 5 MHz, which is 36.94 dB lower than the free-running DFB laser diode.

To verify the wavelength self-adaptivity of the distributed feedback provided by the deformed microcavity, we modified the thermal resistance integrated in the pump (Newport 6100) of the DFB laser diode to tune its wavelength. In a maximum allowable tuning range of 2.25 nm, we measured the linewidth of the DFB laser diode with different external deformed microcavities. As shown in Fig. 5a, the linewidth varies only in a small range for each deformed microcavity, indicating that the deformed microcavity can provide wavelength self-adaptive optical feedback to the lasing cavity. Moreover, the standard deviation of the linewidth decreases with the deformation parameter $\alpha$, as shown in Fig. 5b. For comparison, we also fabricated an add-drop microring resonator with the same radius of 250 μm on the same SOI chip. The resonant spectrum of the add-drop microcavity with a $Q$-factor of $1.21\times10^5$ is shown in Fig. 5c. In the same wavelength tuning range, the linewidth of DFB laser diode with the add-drop microcavity demonstrates strong wavelength dependence, as shown in Fig. 5c, where the narrowest linewidth is 5.125 kHz. The standard deviation of the linewidth is as high as 6.717 kHz with an average linewidth of 25.277 kHz.

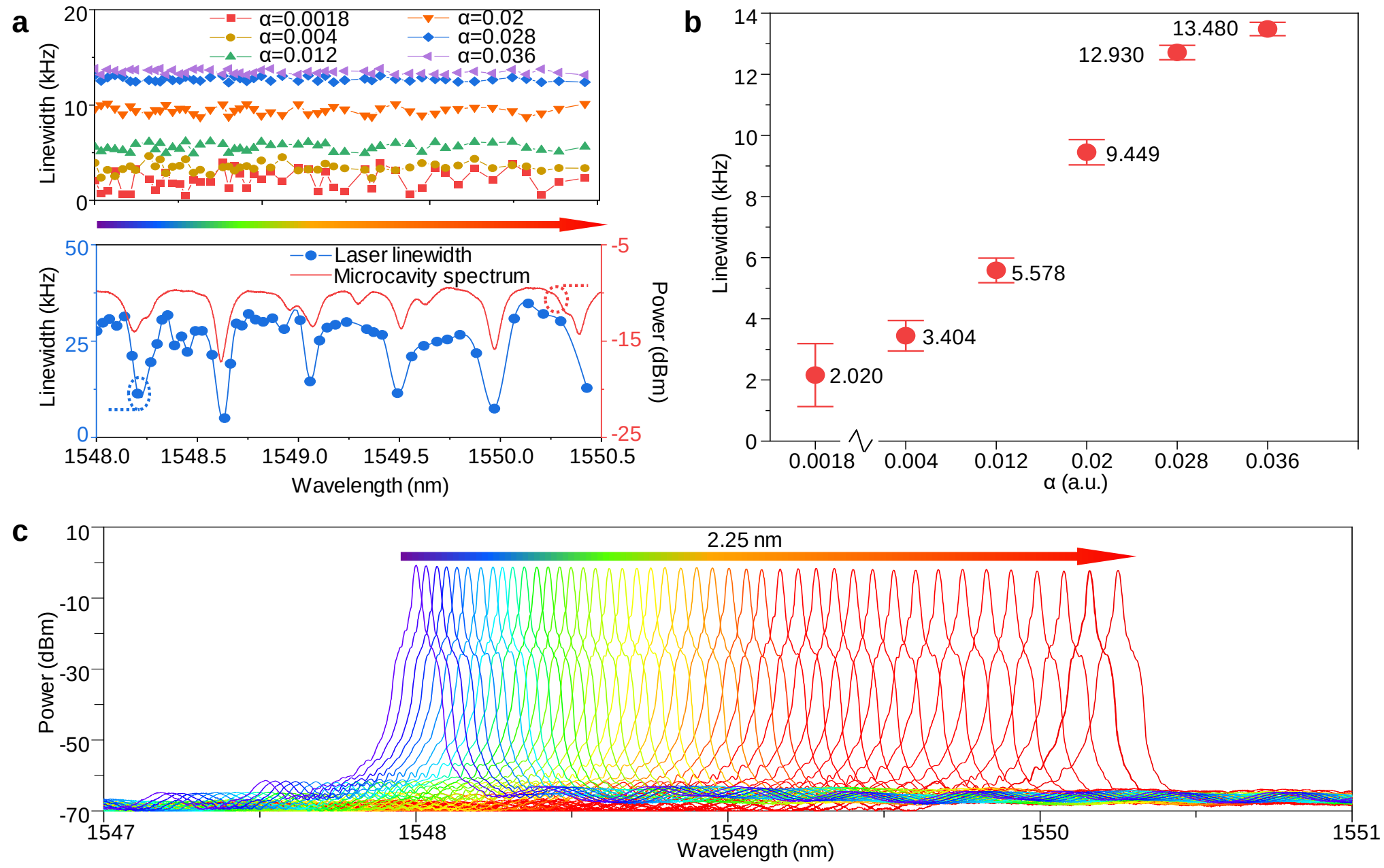

**Fig. 5. Wavelength-adaptivity of the deformed microcavity. a** Laser linewidth versus wavelength under distributed feedback from deformed microcavity with different α . For comparison, resonant feedback from an add-drop microring resonator (with the same diameter and material) is simultaneously measured. **b** Laser linewidth and standard deviation under feedback from deformed microcavities with different α. **c** Thermal tuning of the laser wavelength, with a tuning step of 0.2 kΩ (thermal resistance value).

Finally, we compared the DFB laser diode with deformed microcavity to the DFB laser diode self-injection locked to the add-drop microcavity and a commercially available fiber laser (NKT, Koheras BasiK E15) in terms of linewidth, RIN and frequency noise. As shown in Fig. 6a, the fiber laser demonstrates the narrowest linewidth, which is ~91 Hz. While the RIN of the fiber laser is ~15 dB higher than the two laser diodes, accompanying with a relax oscillation peak at 250 kHz, as shown in

Fig.6b. The frequency noise spectrum in Fig.6c shows that the laser diode with deformed microcavity demonstrate comparable white frequency noise with the fiber laser. In contrast to fiber lasers, semiconductor lasers usually demonstrate low RIN but high frequency noise. However, the DFB laser diode with external distributed feedback demonstrates lower RIN and comparable frequency noise with fiber lasers. Such a breakthrough is of great importance for coherent integrated photonics.

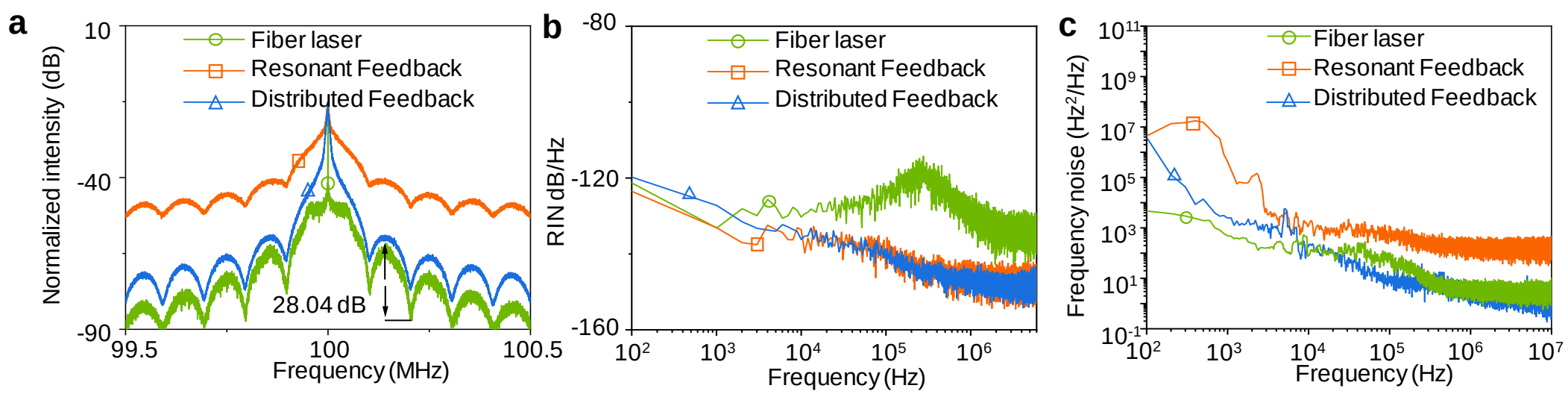

**Fig. 6. Performance comparison of three lasers: laser with distributed feedback, laser with resonant feedback, and commercially available fiber laser. a** Linewidth. **b** Relative intensity noise (RIN). **c** Frequency noise.

# Conclusion

We have demonstrated a hybrid integrated narrow linewidth semiconductor laser with an external deformed microcavity. The frequency noise and RIN are reduced to the level of 2.98 $Hz^2/Hz$ and -148.74 dB/Hz @1 MHz. The wavelength self-adaptive optical feedback provided by the deform microcavity with vortex radius enables similar linewidth narrowing in a wide wavelength range. The work demonstrated here not only addresses the wavelength-dependence induced limitations in conventional microresonators, but also significantly simplifies the wavelength calibration and control processes. The advancement positions the solution as a competitive candidate for integrated narrow linewidth lasers, especially for multi-wavelength laser array or tunable lasers. Furthermore, the concept of backscattering devices without wavelength-selectivity enables the integration of numerous optical technologies that are otherwise challenging to implement on chip-based platforms, such as advanced laser systems including random lasers and chaotic lasers, and optical sensing systems like backscattering based OFDR and Brillouin optical time domain reflectometry (BOTDR). In summary, the deformed cavity and the concept of collecting backscattered signal without wavelength-selectivity demonstrated here provide a new way for broadening the applications of integrated photonics.